\documentclass{PoS}

\usepackage{amsmath}

\usepackage{tikz}
\usetikzlibrary{decorations.markings}
\usetikzlibrary{shapes.geometric}
\usetikzlibrary{patterns}

\usepackage{xcolor}
\usepackage{amssymb}
\usepackage{comment}

\title{Conformal Symmetry and Feynman Integrals}

\ShortTitle{Conformal Symmetry and Feynman Integrals}

\author{\speaker{Simone Zoia} \thanks{In collaboration with Dmitry Chicherin, Johannes M. Henn and Emery Sokatchev.}\\
        Johannes Gutenberg-Universit\"at Mainz\\
        E-mail: \email{zoia@uni-mainz.de}}

\abstract{We develop a new technology for calculating Feynman integrals by solving differential equations derived from conformal symmetry. 
Singularities hidden in the collinear region around an external massless leg may lead to conformal symmetry breaking in otherwise conformally invariant finite loop momentum integrals. 
For an \mbox{$\ell$-loop} integral, this mechanism leads to a set of linear \mbox{$2$nd-order} differential equations with a non-homogeneous part. The latter, due to the contact nature of the anomaly in momentum space, 
is determined by \mbox{$(\ell-1)$-loop} information. Solving such differential equations in general is an open problem. 
In the case of $5$-particle amplitudes up to two loops, the function space is known, and we can thus follow a bootstrap approach to write down the solution. 
As a first application of this method, we bootstrap the 6D \mbox{penta-box} integral.}

\FullConference{Loops and Legs in Quantum Field Theory (LL2018)\\
		29 April 2018 - 04 May 2018\\
		St. Goar, Germany}

\begin{document}

\section{Introduction}
\label{sec:Introduction}

Whenever masses can be neglected, the Poincar\'e group can be extended to the conformal group by the addition of dilatations and conformal boosts (or special conformal transformations).
An extensive study of conformal symmetry in position space has been fostered by the fact that the generators of these transformations are only $1^{\text{st}}$-order differential operators there.
Indeed, standard methods for calculating correlation functions using conformal symmetry date back to the '70s.

Now we want to investigate its implications for scattering amplitudes. Conformal symmetry has thus to be considered directly in momentum space, as the presence of on-shell external momenta prevents us
from Fourier transforming from position to momentum space. The realization of conformal symmetry in momentum space is however significantly more intricate, because the generator of conformal boosts becomes a $2^{\text{nd}}$-order 
operator 
\begin{align}
\label{eq:K}
 K_{\mu} = \sum_{i=1}^n \left[ - p_{i \mu} \Box_{p_i} + 2 p_i^{\nu} \frac{\partial}{\partial p_i^{\nu}}  \frac{\partial}{\partial p_i^{\mu}} + 2 (D-\Delta_i) \frac{\partial}{\partial p_i^{\mu}}  \right],
\end{align}
where $p_i$ are the on-shell massless external momenta, $D$ is the spacetime dimension, $\Delta_i$ is the conformal dimension of the external particle with momentum $p_i$, 
and $\Box_{p_i} = \frac{\partial}{\partial p_i} \cdot \frac{\partial}{\partial p_i}$.

External on-shell massless legs, moreover, are also responsible of a subtle conformal symmetry breaking mechanism. Consider finite Feynman integrals built out of a classically conformally invariant Lagrangian. 
Given the absence of divergences, one might na\"ively expect that such objects are exactly conformally invariant. For this reason, we call them ``na\"ively conformal'' Feynman integrals. Yet, singularities hidden in the 
collinear region around an external massless leg may lead to the breakdown of conformal symmetry. Such collinear anomaly has an intrinsically holomorphic nature, and has long been known at tree-level~\cite{HolomorphicAnomaly}. Its implications for finite loop integrals, however, were only recently taken into account~\cite{Chicherin:2017bxc, Chicherin:2018ubl}.

In section~\ref{sec:AnomalousConformalWardIdentities} we will show how the anomalous conformal Ward identities resulting from this mechanism naturally suggest a new technique for calculating finite loop integrals. We will
then see in section~\ref{sec:6Dbox} that the direct solution of such equations is possible only in simple cases, such as the 6D 1-mass box integral, whereas a bootstrap strategy, described in section~\ref{sec:TheBootstrapStrategy}, 
is more convenient for 5-particle scattering. In section~\ref{sec:6Dpentabox} we will present how the symbol of the 6D penta-box integral can be calculated through the proposed technique, and give an outlook in section~\ref{sec:Outlook}.


\section{Anomalous conformal Ward identities}
\label{sec:AnomalousConformalWardIdentities}
 
\begin{figure}[t!]
\begin{center}
\begin{tikzpicture}[decoration={markings, mark=at position 0.6 with {\arrow{>}}}, scale=1.2]
\draw [thick,postaction={decorate}] (0,0) -- (0.6,0);
\draw [thick,postaction={decorate}] (1.17,0.39) -- (0.6,0);
\draw [thick,postaction={decorate}] (1.17,-0.39) -- (0.6,0);
\draw [pattern=north west lines, pattern color=blue, thick] (1.7,0) ellipse (0.7cm and 0.6cm);
\draw [thick,postaction={decorate}] (2.23,-0.39) -- (2.6,-0.6);
\draw [thick,postaction={decorate}] (2.23,0.39) -- (2.6,0.6);
\draw [thick,postaction={decorate}] (2.35,-0.24) -- (2.6,-0.3);
\draw [thick,postaction={decorate}] (2.35,0.24) -- (2.6,0.3);
\node[text width=0.3cm] at (-0.15,-0.03) {$p$};
\node[text width=0.3cm] at (0.87,-0.44) {$q$};
\node[text width=1.8cm] at (0.73,0.448) {$-p-q$};
\node[text width=0.2cm] at (2.58,0) {\small{...} };
\draw [thick,blue, dashed] (0.65,0) circle[radius=0.2];

\end{tikzpicture}
\end{center}

\caption{Example of an on-shell corner in 6D massless scalar $\phi^3$ theory: an external leg with on-shell massless momentum $p$ enters the diagram and meets two internal lines with off-shell momenta $q$ and $-p-q$. All momenta
are considered as entering the vertex.}
\label{fig:OnshellCorner}

\end{figure}
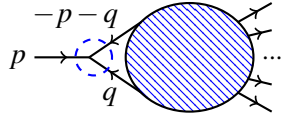

Let us consider for simplicity massless scalar $\phi^3$ theory in $D=6$ dimensions. Acting on each on-shell corner --~see figure~\ref{fig:OnshellCorner}~-- the generator of conformal boosts~\eqref{eq:K} produces a contact anomaly
\begin{align}
\label{eq:KonshellCorner}
 K_{\mu} \frac{1}{q^2 (p+q)^2} = 4 i \pi^3 p_{\mu} \int_0^1 d  \xi \xi (1-\xi) \delta^{(6)}(q+\xi p),
\end{align}
localized on the collinear configuration \mbox{$q \propto p$}~\cite{Chicherin:2017bxc}.

If the diagram considered is tree-level, conformal symmetry is anomalous only for particular configurations of the external momenta, and this effect can thus be neglected for generic kinematics.
If the internal momentum $q$ is flowing in a loop, instead, the Dirac delta on the right-hand side of eq.~\eqref{eq:KonshellCorner} freezes that loop integration, yielding a regular, non-contact anomaly, determined
by lower order information as shown in figure~\ref{fig:KonshellCornerFigure}.

\begin{figure}[h!]

\begin{center}
\begin{tikzpicture}[scale=0.4, decoration={markings, mark=at position 0.6 with {\arrow{>}}}]
      
	\node[text width=1.cm] at (-5.5,0.6){\begin{equation*} K_{\mu} \int d^6 q \end{equation*}};
	\node[text width=5.cm] at (12.5,0.6){\begin{equation*} \propto p_{\mu} \int_0^1 d\xi \xi (1-\xi) \mathcal{I}(q=-\xi p, ...) \end{equation*}};
	\node[text width=1.cm] at (1.5,0) {$\mathcal{I}(q,...)$};
      
	\draw [thick] (2,0) ellipse (2.2cm and 1.7cm);

	\draw [thick,postaction={decorate}] (3.77,-1.) -- (5.8,-1.7);
	\draw [thick,postaction={decorate}] (3.77,1.) -- (5.8,1.7);
	\draw [thick,postaction={decorate}] (4.075,-0.55) -- (5.8,-1.);
	\draw [thick,postaction={decorate}] (4.075,0.55) -- (5.8,1.);
	\node[text width=0.2cm] at (4.8,0) {...};

	\draw [thick,postaction={decorate}] (.22,-1) -- (-1.3,0);
	\draw [thick,postaction={decorate}] (.22,1) -- (-1.3,0);
	\draw [thick,postaction={decorate}] (-2.8,0) -- (-1.3,0);	
	\draw [thick,blue, dashed] (-1.26,0) circle[radius=0.5];
	
	\node[text width=0.3cm] at (-2.4,0.4) {$p$};
	\node[text width=0.3cm] at (-0.82,-1.06) {$q$};

\end{tikzpicture}
\end{center}

\caption{The generator of conformal boosts $K_{\mu}$, acting on the highlighted on-shell corner, produces a contact anomaly, which trades the loop integration 
 for a 1-parameter integration of a lower order object.}
\label{fig:KonshellCornerFigure}

\end{figure}

As a result, the conformal Ward identities for, say, a $\ell$-loop integral $\mathcal{I}^{(\ell)}$ feature an anomaly part $\mathcal{A}_{\mu}^{(\ell-1)}$, obtained from $(\ell-1)$-loop information
\begin{align}
\label{eq:ConformalWardIdentities}
 K_{\mu} \delta^{(6)}\left( p_1+...+p_n  \right) \mathcal{I}^{(\ell)} = \delta^{(6)}\left( p_1+...+p_n \right) \mathcal{A}_{\mu}^{(\ell-1)}.
\end{align}
This set of non-homogeneous linear $2^{\text{nd}}$-order differential equations suggests a method for calculating multi-loop integrals from lower order ones, provided that
such system can indeed be solved for $\mathcal{I}^{(\ell)}$. 
As we shall see in the following sections, this is a highly non-trivial problem, and the direct solution is possible only in simple cases.


\section{A toy example: the 6D 1-mass box}
\label{sec:6Dbox}

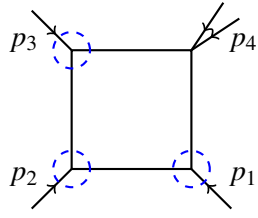
\begin{figure}[t!]
\begin{center}
\begin{tikzpicture}[decoration={markings, mark=at position 0.6 with {\arrow{>}}}, scale=1.05]
\draw [thick] (0,0) rectangle (1.5, 1.5);
\draw [thick, postaction={decorate}] (-0.5,-0.5) -- (0,0);
\draw [thick,postaction={decorate}] (2,-0.5) -- (1.5,0);
\draw [thick,postaction={decorate}] (-0.5,2) -- (0,1.5);
\draw [thick,postaction={decorate}] (1.9,2.1) -- (1.5,1.5) ;
\draw [thick,postaction={decorate}] (2.1,1.9)--(1.5,1.5);
\node[text width=1.cm] at (-0.3,1.6) {$p_3$};
\node[text width=1.cm] at (-0.3,-0.1) {$p_2$};
\node[text width=1.cm] at (2.45,-0.1) {$p_1$};
\node[text width=1.cm] at (2.45,1.6) {$p_4$}; 
\draw [dashed, thick, blue] (1.5,0) circle[radius=0.23];
\draw [dashed, thick, blue] (0,0) circle[radius=0.23];
\draw [dashed, thick, blue] (0,1.5) circle[radius=0.23];
\end{tikzpicture}
\end{center}
\caption{Box integral with a massive leg $p_4^2\neq 0$. Dashed blue lines highlight the on-shell corners.}
\label{fig:1massBox}
\end{figure}

As a first simple application, let us consider the 6D box integral with one massive leg $\mathcal{I}_{1m}$ depicted in figure~\ref{fig:1massBox}.
In this case, the conformal Ward identities~\eqref{eq:ConformalWardIdentities} can be directly solved. This is mainly due to the fact that, once a dimensionful parameter has been factored out, this is just a $2$-variable
problem. Requiring the apparent graph symmetry under exchange of $p_1$ and $p_3$, and the absence of $u$-channel singularities fixes entirely the solution.

However, there is a reason why we do not show here the explicit differential equations: even in such a simple, directly solvable case, they look already rather involved. Indeed, the solution was made possible by 
the following insightful observation, which will play a crucial role also in the more complicated example discussed in section~\ref{sec:6Dpentabox}.

While the $1$-mass box integral is a uniform transcendental weight-$2$ function, the projections of the anomaly along the external momenta $p_i$ exhibit both weight-$0$ and weight-$1$ parts. 
A substantial simplification is achieved by projecting the Ward identities along a vector $q$ such that the anomaly is a weight-0 function 
\begin{align}
 (q\cdot K) \underbrace{\mathcal{I}_{1m}}_{\text{weight-}2} = \underbrace{q \cdot \mathcal{A}}_{\text{weight-}0}.
\end{align}
In other words, the simplest projection of the anomalous conformal Ward identities is the one which features the maximum weight drop.

Anyway, for integrals with kinematics depending on a greater number of variables, a different approach is necessary.


\section{The bootstrap strategy: pentagon functions}
\label{sec:TheBootstrapStrategy}

The bootstrap approach is a general method for solving a broad class of problems, from integrals to differential equations. The basic idea is simple, yet powerful: write down an ansatz
exploiting all the available information about the answer, and impose constraints to fix the free coefficients.
For a finite loop integral $\mathcal{I}$, the ansatz has the general form
\begin{align}
 \mathcal{I}(s) = \sum_{i,j} c_{ij} r_i(s) f_j(s),
\end{align}
where $c_{ij}$ are numeric coefficients to be fixed --~in our case through conformal Ward identities~-- whereas $r_i(s)$ and $f_j(s)$ are respectively algebraic and special functions of the kinematic \mbox{variables $s$}.

While the algebraic functions $r_i(s)$ are controlled by the leading singularities of the integral~\cite{LeadingSingularities}, the class of special functions $f_j(s)$ involved is determined, for
multiple-polylogarithmic functions, by the so-called symbol alphabet~\cite{Goncharov:2010jf}.


\subsection{Basics of symbol technology}
Multiple-polylogarithms can be written as iterated integrals of logarithmic 1-forms~$d \log \alpha_i$, where $\alpha_i$ are algebraic functions of the kinematics. 
Such forms --~usually omitting the logarithm sign~-- are called letters. Together they form the alphabet \mbox{$\mathbb{A} = \{ \alpha_i \}$}. The symbol map, then, associates to a logarithmic iterated integral the tensor product of relevant letters 
\begin{align}
\label{eq:SymbolDefinition}
 S \left[ \int_{\gamma} d \log \alpha_1 \circ  d \log \alpha_2 \circ ... \circ d \log \alpha_w \right] = \alpha_1 \otimes ... \otimes \alpha_w, \qquad \alpha_i \in \mathbb{A} \ \forall i,
\end{align}
where the number of iterated integrations $w$ is called weight of the function.

While differentiation only acts on the last entry of the symbol
\begin{align}
\label{eq:dSymb}
d (\alpha_1 \otimes ... \otimes \alpha_w) = d \log \alpha_w (\alpha_1 \otimes ... \otimes \alpha_{w-1}),
\end{align}
the first entry encodes all the information about the discontinuities 
\begin{align}
\label{eq:DiscSymb}
\text{Disc}\left[ \alpha_1 \otimes ... \otimes \alpha_w \right] = \text{Disc}\left[ \log \alpha_1 \right] (\alpha_2 \otimes ... \otimes \alpha_{w}).
\end{align}

Transcendental constants are mapped to $0$. Therefore, the symbol uniquely fixes the highest functional transcendality part of the function, but misses the lower ones. For instance,
\begin{align*}
 \text{Li}_2(x)+\text{Li}_2(1-x) + \log(x) \log(1-x) = \frac{\pi^2}{6} \Rightarrow S\biggl[ \text{Li}_2(x)+\text{Li}_2(1-x) + \log(x) \log(1-x) \biggr] = 0.
\end{align*}
Some additional work is thus in general necessary to fix those as well.

Furthermore, given an alphabet \mbox{$\mathbb{A} = \{ \alpha_i \}$}, an arbitrary symbol 
\begin{align}
\omega = \sum_{i_1,...,i_w} c_{i_1...i_w} \alpha_{i_1} \otimes ... \otimes \alpha_{i_w}, \qquad c_{i_1...i_w} \in \mathbb{Q}
\end{align}
does not lie in the image of the symbol map in general. A necessary and sufficient condition for the symbol $\omega$ to be integrable to a function is that
\begin{align}
\label{eq:IntegrabilityConditions}
 \sum\limits_{i_1,...,i_w} c_{i_1...i_w} \left( d \log \alpha_{i_a} \wedge d \log \alpha_{i_{a+1}}  \right) \alpha_{i_1} \otimes ... \otimes \hat{\alpha}_{i_a} \otimes \hat{\alpha}_{i_{a+1}} \otimes
  ... \otimes \alpha_{i_w}  = 0, \qquad \forall 1\le a < w,
\end{align}
where hats denote letters which must be omitted, and $\bigwedge$ is the usual exterior product. 

Once the symbol is known, it might be desirable to upgrade it to a function which can be evaluated numerically. 
Provided that a rational parametrization of the kinematics is available, this procedure is completely algorithmic.

\subsection{Pentagon functions}
The scattering of five massless particles is described by five independent variables, which can be conveniently chosen to be the adjacent scalar products $v_i = 2 p_i \cdot p_{i+1}$, where the
indices are cyclic \mbox{mod $5$}.
The alphabet describing on-shell 5-particle scattering amplitudes is known up to two loops~\mbox{\cite{Gehrmann:2015bfy, Chicherin:2017dob}}, and counts 31 letters
\begin{align}
 &  \alpha_i = v_i, & &  \alpha_{5+i} = v_{i+2} + v_{i+3}, \nonumber \\
 &  \alpha_{10+i} = v_i - v_{i+3}, & & \alpha_{15+i} = v_i + v_{i+1}-v_{i+3},   \nonumber \\ 
 &   \alpha_{20+i} = v_{i+2}+v_{i+3}-v_{i}-v_{i+1},& & \alpha_{25+i} = \frac{a_i-\sqrt{\Delta}}{a_i+\sqrt{\Delta}},  \nonumber \\ 
 & \alpha_{31} = \sqrt{\Delta}, & & 
\end{align}
where $i$ runs from $1$ to $5$, $\Delta$ is the Gram determinant $\Delta = \text{det}\left(2 p_i \cdot p_j \right)$, and
\begin{align}
a_i = v_{i} v_{i+1} - v_{i+1} v_{i+2} + v_{i+2} v_{i+3} - v_{i+3} v_{i+4} - v_{i+4} v_i.
\end{align}

Of these letters, only 26 are relevant for planar diagrams
\begin{align}
 \mathbb{A}_{\text{P}} = \{\alpha_1,...,\alpha_{20} \} \bigcup \{ \alpha_{26},..., \alpha_{31} \}.
\end{align}


Recalling from eq.~\eqref{eq:DiscSymb} that the information about the discontinuities of a function is contained in the first entries of its symbol, general considerations about the locations of physical branch points lead to a \textit{first entry condition}: 
only the 10 letters $\{\alpha_1,...,\alpha_5\} \bigcup \{\alpha_{16},...,\alpha_{20}\}$ are allowed as first entries in the symbol of a non-planar function; in the planar case, this list reduces to $\{\alpha_1,...,\alpha_5\}$.

A \textit{second entry condition} is conjectured in ref.~\cite{Chicherin:2017dob}. Although it significantly constrains the number of allowed functions, as can be seen in 
table~\ref{tab:IntegrableFunctions}, we are not going to take it into account.
\begin{table}[t!]
\begin{center}
\begin{tabular}{|l|c|c|c|c|c|c|}
    \hline
    Weight & 1 & 2 & 3 & 4 & 5 & 6 \\
    \hline
    $1^{\text{st}}$ entry condition & $5$ & $25$ & $126$ & $651$ & $3436$ &  $18426$ \\
    $2^{\text{nd}}$ entry condition & $5$ & $20$ & $81$ & $346$ &  $1551$ & $7201$ \\
    \hline
  \end{tabular}
\end{center}
\caption{Number of integrable planar pentagon symbols up to weight 6 with first and second entry condition.}
\label{tab:IntegrableFunctions}
\end{table}


\section{A non-trivial example: the 6D penta-box}
\label{sec:6Dpentabox}

\begin{figure}
\begin{center}

\begin{tikzpicture}[decoration={markings, mark=at position 0.6 with {\arrow{>}}}, scale=1.15]
\begin{scope}[yscale=1,xscale=-1]

    \foreach \x in {0,72,...,288} { \draw (\x:1 cm) -- (\x + 72:1 cm); };
    \draw (-0.809017,0.587785)--(-1.98459 ,0.587785 )--(-1.98459,-0.587785 )--(-0.809017,-0.587785) ;

    \draw[postaction={decorate}]  (1.8,0)--(1,0);
    \draw[postaction={decorate}]  (0.556231,1.7119)--(0.309017, 0.951057);
    \draw[postaction={decorate}]  (0.556231,-1.7119)--(0.309017,-0.951057);
    \draw[postaction={decorate}]  (-2.55028 ,1.15347 )--(-1.98459 ,0.587785 );
    \draw[postaction={decorate}]  (-2.55028 ,-1.15347 )--(-1.98459 , -0.587785 );
    
    \node[text width=1cm] at (+0.45, -1.45) {$p_1$};    
    \node[text width=1cm] at (+0.45, 1.45) {$p_3$};
    \node[text width=1cm] at (1.4, 0.2) {$p_2$};
    \node[text width=1cm] at (-2.45 ,1.15347 ) {$p_4$};
    \node[text width=1cm] at (-2.45 ,-1.15347 ) {$p_5$};
    
\end{scope}
\end{tikzpicture}
\end{center}
\caption{Penta-box integral. All external momenta are on-shell massless and entering.}
\label{fig:Pentabox}
\end{figure}
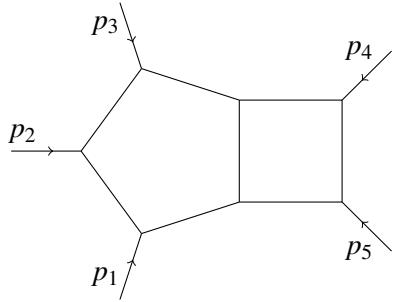

As first non-trivial application of the conformal symmetry method, we tackle the calculation of the 6D penta-box integral $\mathcal{I}_5$ depicted in figure~\ref{fig:Pentabox}. 

A 2-loop planar integral in $D=6$ in general can have at most transcendental weight $6$, resulting in $18426$ integrable symbols to be taken into account
(see table~\ref{tab:IntegrableFunctions}). Quite remarkably, this number can be significantly cut down by conformal symmetry even before calculating the anomaly. 

To see this, note that the 6D penta-box has only one leading singularity, $1/\sqrt{\Delta}$, which is conformally invariant, and that for a generic weight-$W$ symbol $\alpha_{i_1} \otimes ... \otimes \alpha_{i_W}$
\begin{align}
\label{eq:qKsymb}
  \left[ (q \cdot K) \frac{\alpha_{i_1} \otimes ... \otimes \alpha_{i_W}}{\sqrt{\Delta}} \right] = \left[ (q \cdot K) \frac{\log \alpha_{i_W}}{\sqrt{\Delta}}\right] 
 (\alpha_{i_1} \otimes ... \otimes \alpha_{i_{W-1}}) + \text{weight-}(W-2),
\end{align}
where $K_{\mu}$ is given by eq.~\eqref{eq:K}, and $q$ is a generic vector.

Here comes into play the lesson we learnt from the $1$-mass box example in section~\ref{sec:6Dbox}. If we can find a vector $q$ along which the weight drop is maximal
\begin{align}
 (q\cdot K) \underbrace{\mathcal{I}_{5}}_{\text{weight-}W} = \underbrace{q \cdot \mathcal{A}}_{\text{weight-}(W-2)},
\end{align}
the whole weight-$(W-1)$ part of $(q\cdot K)\mathcal{I}_{5}$ would vanish. Via eq.~\eqref{eq:qKsymb}, this would lead to a \textit{last entry condition}: only the letters $\alpha_i \in \mathbb{A}_{\text{P}}$ fulfilling
\begin{align}
\displaystyle (q \cdot K) \left[ \sum\limits_{i} c_i \frac{\log \alpha_{i}}{\sqrt{\Delta}} \right] = 0
\end{align}
are allowed as last entries in the symbol of the 6D penta-box integral.

Such a projection does indeed exist. If we consider a vector $q^* \perp p_2, p_4, p_5$, we find that
\begin{align}
\label{eq:qAnomaly}
 \left(q^* \cdot \mathcal{A}\right) = \frac{(\text{weight-}3)}{s_{12} s_{14} (s_{23}-s_{45})}  + (1\leftrightarrow 3, 4 \leftrightarrow 5), 
\end{align}
where $s_{ij} = 2 p_i \cdot p_j$. This result has two important implications. 

First, $\mathcal{I}_5$ cannot have weight 6: the anomaly contains a weight-3 part, and a \mbox{$2^{\text{nd}}$-order} operator
can reduce the weight at most by 2. Therefore, and since there is only one leading singularity, we can assume that $\mathcal{I}_5$ is a uniform weight-5 function, and our ansatz for the symbol of $\mathcal{I}_5$ is
\begin{align}
\label{eq:Ansatz}
 S\left[ \mathcal{I}_5 \right] = \frac{1}{\sqrt{\Delta}} \sum_{i_1,...,i_5} c_{i_1...i_5} \left(\alpha_{i_1} \otimes ... \otimes \alpha_{i_5} \right), \qquad \alpha_k \in \mathbb{A}_{\text{P}} \ \forall k.
\end{align}

Second, only 11 out of the 26 letters of $\mathbb{A}_{\text{P}}$ are allowed as last entries. Supplying this constraint with information about the symmetries of the integral, namely that
\begin{itemize}
 \item $\mathcal{I}_5$ is even under complex conjugation and, since $\sqrt{\Delta}$ is odd, the symbol has to be odd,
 \item the diagram, see figure~\ref{fig:Pentabox}, is symmetric under the exchange  $\{p_1 \leftrightarrow p_3, p_4 \leftrightarrow p_5 \}$,
\end{itemize}
the number of symbols allowed in the ansatz~\eqref{eq:Ansatz} can be cut down to only 33, see table~\ref{tab:Constraints}.
\begin{table}[h!]
\centering
\begin{tabular}{|c|c|}
\hline
 Constraints & weight-5 integrable symbols ($\mathbb{A}_\text{P}$)    \\
\hline
\hline
  $1$st entry condition & 3436 \\
 odd under complex conjugation & 161 \\
 last entry condition & 59  \\
 exchange symmetry & 33  \\
\hline
\end{tabular}
\caption{Number of weight-5 integrable symbols from the planar alphabet $\mathbb{A}_{\text{P}}$ fulfilling gradual constraints.}
\label{tab:Constraints}
\end{table}

Finally, plugging the ansatz into the explicit anomalous conformal Ward identities
fixes all the remaining coefficients, yielding an expression for the symbol of the 6D penta-box which agrees with both the known result, and the conjectured $2^{\text{nd}}$-entry condition of ref.~\cite{Chicherin:2017dob}.

Not only all the coefficients are fixed, but just one projection of the conformal Ward identities --~along the vector $q^* \perp p_2, p_4, p_5$~-- is sufficient to do so. On top of that, such projection is actually the simplest, 
as $q^*$ annihilates the contributions to the anomaly involving the pentagon sub-integral. Instead, only the contributions from the on-shell corners $p_1$ and $p_3$ survive, which can be obtained one from the
other by exploiting the graph symmetry, and only involve the 1-mass box. Considering that the symbol of the 6D penta-box counts around 10000 terms, 
this successful application gives very solid evidence of the power of conformal symmetry in constraining finite loop integrals.


\section{Outlook}
\label{sec:Outlook}
Since the kinematics can be parametrized so as to rationalize the pentagon alphabet, the symbol of the 6D penta-box can be algorithmically upgraded to function. However, the lower functional transcendentality piece is still missing.
In order to fix it, we plan to upgrade the bootstrap approach to function level by writing an ansatz for it directly
in terms of Goncharov polylogarithms. Which transcendental constants should be accommodated in the ansatz is however an open problem.

Although this talk was only concerned with $6D$ $\phi^3$ theory, the mechanism of conformal symmetry breaking is more
general. Indeed, the anomalies of classically conformal $4D$ theories have been derived in ref.~\cite{Chicherin:2017bxc} as well. Finite 4D loop integrals are
currently under investigation.

The same mechanism has also been shown to lead to super-conformal symmetry breaking in
the Wess-Zumino model of $\mathcal{N} = 1$ massless supersymmetric matter~\cite{Chicherin:2018ubl}. However, the generator of super-conformal boosts is $1^{\text{st}}$-order:
 the (super)conformal Ward identities are thus easier, and more powerful. 
Indeed, their direct solution was possible for a non-planar 2-loop 5-particle integral. We therefore expect that this technique
will enable the calculation of even more complicated integrals.


\acknowledgments{I wish to thank Maximilian Stahlhofen for his careful reading of the manuscript, and Tiziano Peraro for providing an independent cross-check of the integrable symbols of the
pentagon alphabet. This project has received funding from the European
Research Council (ERC) under the European Union's Horizon 2020 research and innovation
programme (grant agreement No~ 725110), `Novel structures in scattering amplitudes'. This work was supported by a GFK fellowship.}


\end{document}